\documentclass{PoS}

\title{The GRANIT project: Status and Perspectives}

\ShortTitle{The GRANIT project}
        
\author{Stephan Bae\ss{}ler,$^a$ Alexei Gagarski,$^b$  Ludmilla Grigorieva,$^b$ Michael Kreuz,$^d$ Fabrice Naraghi,$^c$ Valery
Nesvizhevsky,$^d$ Guillaume Pignol,$^c$ Konstantin Protassov,$^c$ \speaker{Dominique Rebreyend}, $^c$ Francis Vezzu,$^c$ Alexei
Voronin$^e$\\
\llap{$^a$} University of Virginia, Charlottesville, VA-22904, USA\\
and Oak Ridge National Laboratory, Oak Ridge, TN-37831, USA\\
\llap{$^b$} PNPI, Orlova Roscha, Gatchina, Leningrad Reg. 188350, Russia\\
\llap{$^c$} LPSC, UJF, CNRS/IN2P3, INPG, 53 rue des Martyrs, Grenoble F-38026, France\\
\llap{$^d$} ILL, 6 rue Jules Horowitz, Grenoble F-38042, France\\
\llap{$^e$} Lebedev Institute, 53 Leninsky Pr., Moscow 119991, Russia\\

E-mail: \email{baessler@virginia.edu}, \email{gagarski@pnpi.spb.ru}, \email{l\_grigoryeva@mail.ru}, \email{kreuz@ill.eu},
\email{naraghi@lpsc.in2p3.fr}, \email{rebreyend@lpsc.in2p3.fr},\email{nesvizhevsky@ill.eu},
\email{pignol@lpsc.in2p3.fr},\email{protasov@lpsc.in2p3.fr}, \email{vezzu@lpsc.in2p3.fr},\email{dr.a.voronin@gmail.com}}

\abstract{The GRANIT project is the follow-up of the pioneering experiments that first observed the quantum states of neutrons
trapped in the earth's gravitational field at the Institute Laue Langevin (ILL). Due to the weakness of the gravitational force,
these quantum states exhibit most unusual properties: peV energies and spatial extensions of order 10~$\mu$m.  Whereas the first
series of observations aimed at measuring the properties of the wave functions, the GRANIT experiment will induce resonant
transitions between states thus accessing to spectroscopic measurements. After a brief reminder of achieved results, the principle
and the status of the experiment, presently under commissioning at the ILL, will be given. In the second part, we will discuss the
potential of GRANIT to search for new physics, in particular to a modified Newton law in the micrometer range. }

\FullConference{The 2011 Europhysics Conference on High Energy Physics-HEP 2011,\\
		July 21-27, 2011\\
		Grenoble, Rh\^one-Alpes France}

\begin{document}

\section{The neutronic quantum bouncer}

Neutrons bouncing off a horizontal mirror under the sole gravitational attraction of the earth may sound unexpected or even
impossible for a high energy physicist. Indeed, neutrons being neutral objects of extremely small dimension, they are known to be
very penetrating particles. Moreover, gravity is by far the weakest of all forces and is usually totally negligible in experiments
dealing with microscopic bodies. This very phenomenon has nevertheless been observed about 10 years ago in a pioneering experiment
at the Institut Laue Langevin (ILL) \cite{nature}, thanks to the use of very low energy neutrons. These neutrons, known as
Ultra-Cold-Neutrons (UCN), have velocities of a few m/s, equivalent to energies in the 100~neV range, and have the extraordinary
property of being reflected at any incidence angle by some materials. As a consequence, they can be stored in containers and are
the privileged tools to perform particle physics experiments at low energy, most notably measurements of the neutron lifetime and
of its electric dipole moment.

\vspace{-4cm}

\begin{figure}[h]
\centering
\includegraphics[width=0.5\linewidth,angle=0]{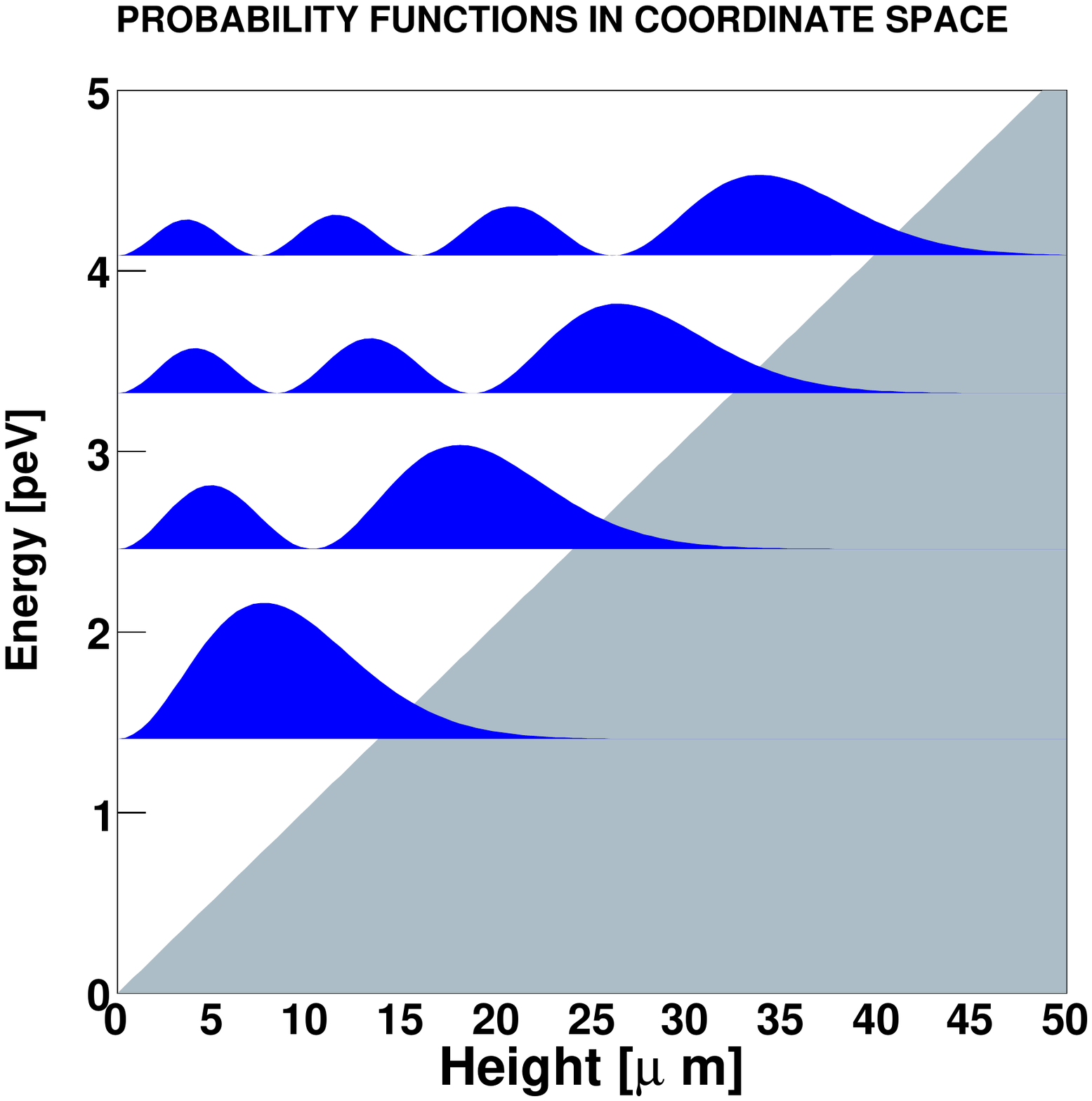}
\caption{Energies and density probability functions of the first 4 quantum states. 
}
\label{waveFunctions}
\end{figure}

In the language of quantum mechanics, neutrons are trapped in a potential well -- formed by a nearly infinite wall due to the
repulsion of the mirror \footnote{Energy of the quantum levels of interest is $\sim$10$^{-12}$ eV to be compared to the 90~neV
repulsive (Fermi) potential of the quartz mirror} on the one hand and the usual linear $m g z$ attractive potential of the earth
on the other hand -- and their energy is therefore discretized. Hence the name neutronic quantum bouncer. 

Fig.\ref{waveFunctions} shows the energy and the probability density function of the first 4 quantum states. Because of the
weakness of the gravitational interaction, these states have outstanding properties: energies are in the peV -- 10$^{-12}$ eV ! --
range and the typical size of the wave-functions is of order 10$\mu$m, i.e of macroscopic extension.

The very first experiment measured the flux of UCN after a slit of varying height, whose upper half was an absorber. In agreement
with the vertical extension of the ground state, UCN could be detected after the slit only when its dimension would be larger than
about 10~$\mu$m. Other measurements were also performed but all based on the neutron vertical distribution i.e. testing only the
wave-function of this system. 

This series of pioneering experiments opened up a new field of research, giving one of the very few possibilities to study
gravitation in the context of quantum mechanics \cite{NOVA,ReviewQuantumStates}. Numerous ideas have been proposed to exploit the
high sensitivity of this system, in particular the possibility to constrain a short range fifth interaction. However, the
precision reached on the parameters of this system, and correspondingly the sensitivity to new physics, was limited to a few
percent and did not allow to improve the current limits \cite{Nes,Baessler09}. 

\section{The GRANIT project}

GRANIT aims at inducing transitions between the quantum states of the neutronic quantum bouncer in order to measure the transition
energies. By doing so, we will access to the high precision of spectroscopic measurements and will increase our sensitivity to
perturbations by orders of magnitude. 

\begin{figure}[h]
\centering
\includegraphics[width=0.7\linewidth,angle=0]{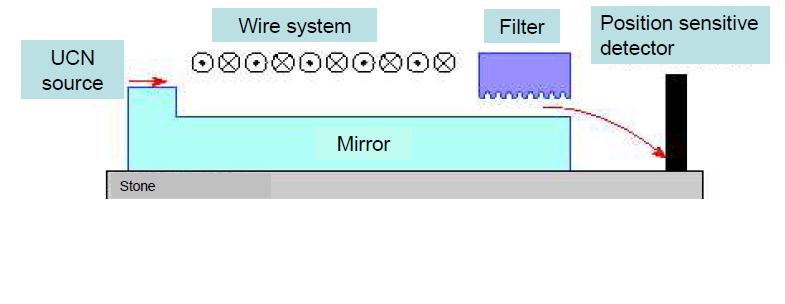}
\vspace{-1cm}

\caption{Principle of the set-up to measure resonant transitions induced by an oscillating magnetic field
gradient in the flow-through mode.}
\label{setup}
\end{figure}

A scheme of the planned initial set-up is displayed in Fig.\ref{setup}. From left to right, the UCN will first be prepared in the
third state by a step of appropriate height. They will then bounce between a mirror and a wire system whose function is to induce
a Raby-like transition to the ground state. The wires are fed by steady currents and produce a spatially periodic magnetic field
gradient. In their rest frame, neutrons will feel an oscillating gradient with a frequency depending on their velocity.  Only
neutrons of a given velocity will be at resonance, i.e. will see a gradient frequency equal to the $1 \rightarrow 3$ transition
frequency, and will then undergo a full transition to the ground state. At the end of the mirror, a filter will select only UCN in
the ground state. The surviving neutrons will then start a free fall until they reach a position-sensitive detector. By measuring
the height of the free fall, one will be able to calculate the horizontal velocity and in fine to access to the $1 \rightarrow 3$
transition frequency.

It should be noted that the gravity-resonance-spectroscopy has been realized by Abele and
collaborators  \cite{Jenke}. Instead of using magnetic gradients, they used mechanical
oscillations of the underlying mirror to induce the transitions.

Most of the equipment is now in place at level C of the ILL. Commissioning has already started and we expect to start real data
taking next Spring. A more detailed description of the apparatus can be found in \cite{Kreuz}.

\section{Constraint on Chameleon models}

In a recent work \cite{cham}, the potential detection of chameleons with the neutronic quantum bouncer has been explored.
Chameleons are scalar fields of the quintessence type and are among the few serious candidates to explain the accelerated
expansion of the Universe. 

\begin{figure}[h]
\centering
\includegraphics[width=0.5\linewidth,angle=90]{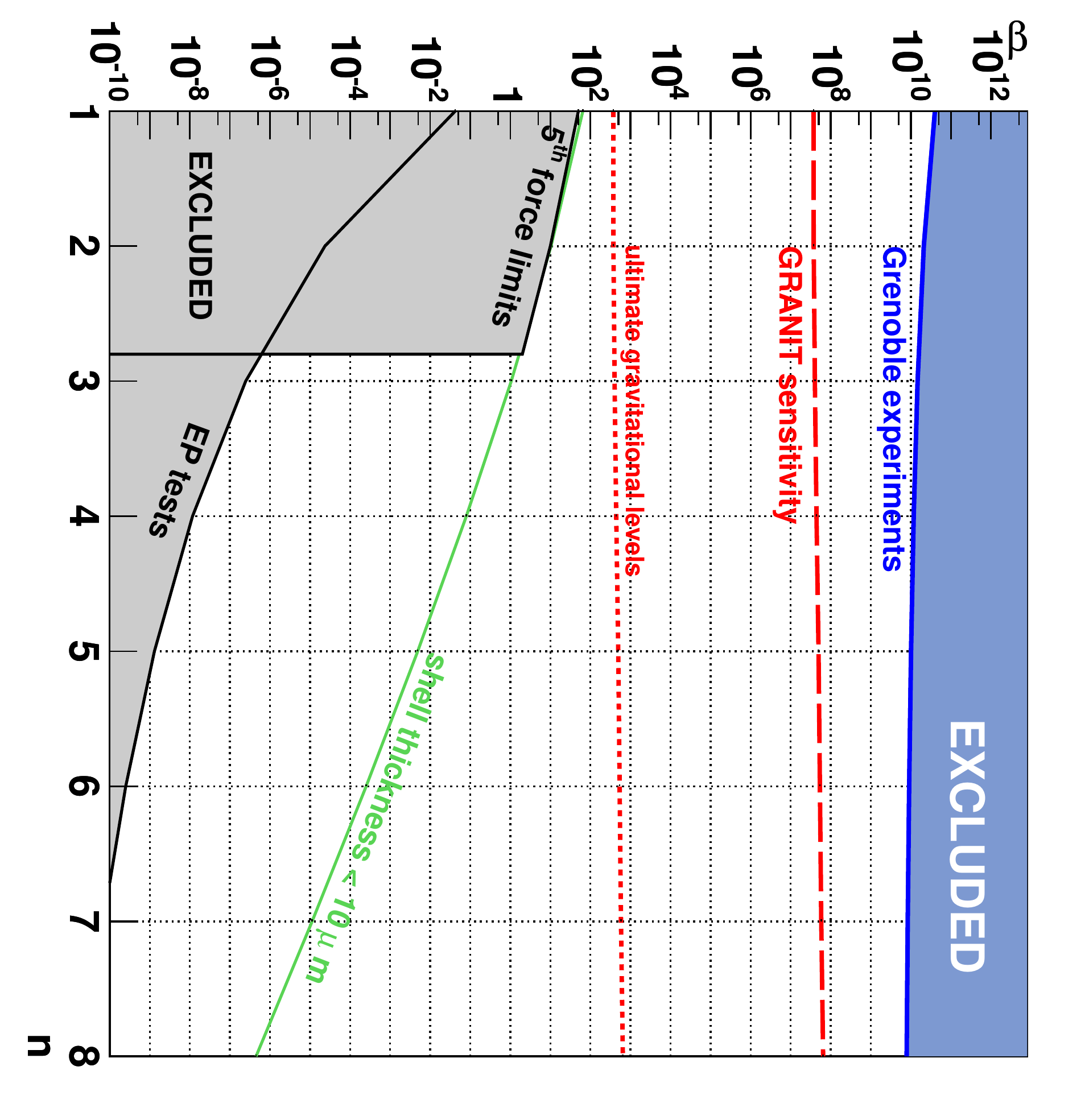}
\caption{The chameleon exclusion plot.
We find that above the  bottom green line the  chameleon field is independent of the coupling,
above the top blue line  chameleons produce a quantum state with a size of 2 microns and finally
above the red dashed line the chameleons shift the $3 \rightarrow 1$ resonance by more than $0.01$~peV. 
We have also drawn the ultimate sensitivity limit at the $10^{-7}$ peV level. 
}
\label{exclusion}
\end{figure}

What makes these scalar fields of particular interest is that they couple to matter, hence acquiring the so-called chameleon
mechanism, by which they could have evaded long range fifth force searches or equivalence principle tests. Because of the coupling
to matter, the field gets trapped inside matter, screening its existence to the outside world. In \cite{cham}, it has been shown
that the existence of a chameleon field would produce an extra attractive term above the mirror and would lead to a modified
gravity potential:

\begin{equation}
\label{pot1}
\Phi(z) = m gz + \beta \frac{m}{M_{\rm Pl}} \phi(z).
\end{equation}

with $\phi(z)$ the chameleon field, $m$ the mass of the neutron, $M_{\rm Pl}$ the Planck mass and $\beta$ the coupling constant.
This extra term would lead to two potentially observable effects: the shrinking of the wave functions of the stationary states and
the shifting of the energy levels. Fig.\ref{exclusion} shows the resulting exclusion plot. 

Macroscopic experiments (Equivalence Principle tests and 5th force limits) exclude the small $\beta$ region. This is due to the
screening effect: when $\beta$ increases, the chameleon thin shell at the surface of the test bodies shrinks and there is no
increase in the force. For the neutronic quantum bouncer, the situation is different. Neutrons have no thin shell effect and, as
seen in equation~\ref{pot1}, the potential felt by the neutron is linear in $\beta$. The very large $\beta$ region is already
excluded by the first series of experiments. For such large coupling, an additional bound state near the surface of the mirror
would appear and would have been detected. In the first stage of the GRANIT experiment, the energy difference $E_3 - E_1 $ is
expected to be measured with an accuracy of 0.01~peV which sets the sensitivity to a chameleon induced shift (dashed line in
fig.~\ref{exclusion}). GRANIT will be therefore able to set competitive constraints in the large coupling region of the chameleon
models parameter space.

In summary, GRANIT is a second generation experiment aiming at investigating the properties of the neutron quantum bouncer, using
the high precision of spectroscopic measurements. We expect to improve the precision of the first series of measurememts --
thereby our sensitivity to new physics -- by at least an order of magnitude within the next 3 years.

\end{document}